\documentclass[a4paper,twoside,11pt]{article}

\usepackage[USenglish]{babel} % francais, polish, spanish, ...
\usepackage[T1]{fontenc}
\usepackage{amsmath}

\usepackage{graphicx}
\usepackage{url}
\usepackage[labelsep=endash, figurename=Figure, tablename=Table, textfont=it]{caption}
\usepackage{subcaption}
\usepackage{float}  
\usepackage{hyperref, wasysym}
\usepackage{multicol}
\usepackage{cleveref}
\usepackage{tabularx}
\usepackage{booktabs}
\usepackage{amsfonts}
\usepackage{graphicx}
\usepackage{graphicx}
\usepackage{subcaption}

\usepackage{fancyhdr}
\usepackage[top=2.5cm, bottom=2.5cm, left=2.5cm, right=2.5cm]{geometry} 
\usepackage[usenames,dvipsnames]{pstricks}
\usepackage[round,comma,authoryear]{natbib}
\usepackage{titling}

\definecolor{darkblue}{rgb}{0,0,0.5}
\hypersetup{colorlinks=true,linkcolor=black,citecolor=darkblue,urlcolor=darkblue} 

\makeatletter
\let\ps@plain\ps@fancy
\makeatother

\begin{document}

%% Title Page %%%%%%%%%%%%%%%%%%%%%%%%%%%%%%%%%%%%%%%%%%%%%%%

\title{\bf Improving simulation-based origin-destination demand calibration using sample segment counts data}
\date{}

\pretitle{\centering\Large}
\posttitle{\par\vspace{5pt}}

\preauthor{\centering}
\author{Arwa Alanqary\protect\thanks{University of California, Berkeley} \thanks{Done while at Google Research}, Chao Zhang\protect\thanks{Google Research}, Yechen Li\protect\footnotemark[3], Neha Arora\protect\footnotemark[3], and Carolina Osorio\protect\footnotemark[3] \thanks{HEC Montr\'eal}}
\postauthor{\par\vspace{-8ex}}
\maketitle

\vspace{-5ex}
\noindent\rule{\textwidth}{0.5pt}\vspace{0cm}

% Header and footer for the first page
\fancypagestyle{firststyle}{
\lhead[]{}
\rhead[]{}
\lfoot[TRISTAN XII Symposium]{TRISTAN XII Symposium}
\cfoot[Okinawa, Japan]{Okinawa, Japan}
\rfoot[June 22 - 27, 2025]{June 22 - 27, 2025}
}
\thispagestyle{firststyle}

% Header and footer for all other pages
\pagestyle{fancy}
\fancyhead{}
\fancyfoot{}
\renewcommand{\headrulewidth}{0pt}
\renewcommand{\footrulewidth}{0pt}
\setlength{\headheight}{15pt}
\lhead{}
\rhead[\thepage]{\thepage}
\lfoot[TRISTAN XII Symposium]{TRISTAN XII Symposium}
\cfoot[Okinawa, Japan]{Okinawa, Japan}
\rfoot[June 22 - 27, 2025]{June 22 - 27, 2025}

\vspace{-12pt}
%%%%%%%%%%%%%%%%%%%%%%%%%%%%%%%%%%%%%%%%%%%%%%%%%%%%%%%%%%%%%
%% CONTENTS - Extended abstract
%%%%%%%%%%%%%%%%%%%%%%%%%%%%%%%%%%%%%%%%%%%%%%%%%%%%%%%%%%%%%
\section{Introduction}
\vspace{-0.3 cm}

High-resolution stochastic traffic simulators are vital tools for the planning and evaluation of urban mobility and transportation infrastructures and services. To fully leverage the benefits of these models, effective calibration is essential. One of the most critical calibration tasks involves identifying origin-destination (OD) demand patterns that accurately replicate observed traffic statistics from field measurements. 
%This task is the topic of an active line of research aimed at developing scalable algorithms that solve the demand calibration problem with high accuracy.
%
Calibrating OD demand presents several challenges: the problem is high-dimensional, stochastic, non-linear, and underdetermined. A variety of optimization algorithms have been adapted for demand calibration, including simultaneous perturbation stochastic approximation (SPSA) \cite{ben2012dynamic}, genetic algorithms \cite{chiappone2016traffic}, and metamodel simulation-based optimization (SO) approaches \cite{osorio2019high}. 

Traditionally, segment counts collected from loop detectors have been the primary data used for demand calibration. Advancements in traffic surveillance technologies have introduced new data sources that enhance the demand calibration process. These include mobile phone call data \cite{iqbal2014development}, Bluetooth data \cite{barcelo2012dynamic}, and license plate recognition data \cite{mo2020estimating}. Such data can yield various statistics about traffic flow — including counts, speeds, and travel times — on segment, path, or sub-path levels. The availability of these high-resolution datasets offers significant potential for improving the quality and scalability of OD demand calibration, providing extensive coverage without the need for often prohibitively expensive sensor infrastructure.
Despite their advantages, these data sources have limitations. Although they can provide accurate speed and travel time statistics, the count data they provide can only be interpreted as samples rather than complete measurements of traffic flow, mainly due to limited penetration.

In this paper, we introduce a novel method for demand calibration that integrates path and segment-level traffic statistics from high-resolution, wide-coverage trajectory data. 
%even in the absence of known penetration rates. 
Our methodology builds upon the existing metamodel SO framework \cite{osorio2019high} and recent advancements that utilize path travel times as ground truth data \cite{chao2019traffic}. The segment-level sample counts are integrated into the optimization problem as a regularization term despite unknown penetration rates, penalizing large deviations from the observed count distribution. This allows us to better capture network flow patterns and thus improve the accuracy of the calibrated demand. 
%particularly in OD pairs with low demand and congestion where travel time data alone is insufficient. 
Importantly, our method preserves the scalability and sample efficiency of the original formulation. We evaluate the effectiveness of this approach on a large-scale use case of Seattle highway network across a range of traffic scenarios. 

\vspace{-0.4 cm}
\section{Methodology}
%In this section, we give formal statement of the OD demand calibration problem and detail the proposed methodology. 
\vspace{-0.3 cm}
\subsection{Problem formulation} 
\vspace{-0.3 cm}

 We focus on the \textit{static} demand calibration problem, which aims to identify an OD demand matrix that reflects the observed traffic conditions aggregated over a time interval of interest. We utilize two types of field measurements: (1) the path-level travel time between OD pairs, and (2) a sample of the vehicular count on the segments. This unique combination of data is well-motivated in various contexts. For example, navigation-based field measurements provide accurate path-level travel time estimates and aggregated segment-level counts based on the fraction of commuters that use the navigation service. Both types of measurements have a wide spatial coverage.  

To formulate the calibration problem, let $\mathcal{Z}$ denote the set of OD pairs and $d \in \mathbb{R}_{+}^{|\mathcal{Z}|}$ be the demand vector (flattened demand matrix). We denote by $\mathcal{P}$ and $\mathcal{I}$ the set of paths and segments in the network with field measurements, respectively. Let $y^{\text{GT}}_p$ be the measured travel time on path $p \in \mathcal{P}$ and $x_i^{\text{GT}}$ be the sample counts on segment $i\in \mathcal{I}$. 
We formulate the OD demand calibration problem as the following optimization problem:
\vspace{-10pt}
\begin{align}
    \min_{0 \leq d \leq d_{max}} F(d) = \frac{w_1}{|\mathcal{P}|} \sum_{p\in \mathcal{P}} \left( y_p^{\text{GT}} - \mathbb{E}[Y_p(d; u)] \right)^2 + \frac{w_2}{|\mathcal{I}|} \sum_{i\in \mathcal{I}} \left( \frac{\mathbb{E}[X_i(d; u)]}{x_i^{\text{GT}}} - \frac{1}{\mathcal{|I|}}\sum_{j \in \mathcal{I}} \frac{\mathbb{E}[X_j(d; u)]}{x_j^{\text{GT}}} \right)^2, 
\label{eq_so_problem}
\end{align}
where $\mathbb{E}[Y_p(d; u)]$ and $\mathbb{E}[X_i(d; u)]$ are the expected travel time on path $p$ and expected flow of segment $i$, respectively, defined by the stochastic traffic simulator with exogenous parameter vector $u$. 
The first term in the objective function captures the distance between the travel times from the field measurements and those estimated by the simulator. 
The second term acts as a regularization term, penalizing large variances in the ratio between simulated and observed segment counts. This ensures the solution's physical plausibility and preserves the structure of the observed count data by matching their distributions.
%The second term serves as a regularization term to ensure the physical plausibility of the solution. This is done by penalizing large variances in the ratio between the simulated and observed (sample) segment counts. This helps preserve the structure in the observed sample segment counts data by matching their distributions. 
%
While this formulation is motivated by the assumption that the penetration rates of the sample segment counts are constant across the network, we use it as a regularization term rather than a hard constraint to tolerate the inherent variability in the penetration rates across different segments. 
%
%By proper tuning of the two weight parameters $w_1$ and $w_2$, we aim to find a demand vector that well replicate the travel time measurements, while producing segment counts with spatial distributions similar to the observed sample counts. 
\vspace{-0.3 cm}
\subsection{Metamodel algorithm} 
\vspace{-0.3 cm}
The problem formulated in \cref{eq_so_problem} is a high-dimensional SO problem (with OD variables in the order of hundreds to thousands). It also involves expensive evaluation of a black-box stochastic traffic simulator that is generally non-convex and non-differentiable. To address these challenges, we need sample-efficient algorithms that can reduce the needed number of function evaluations. 
%by guiding the identification of good candidate solutions. 
% %
% For this we resort to the class of metamodel algorithms, which have been proven effective for addressing these challenges in SO problem in general and the formulated demand calibration problem in specific in a salable and sample efficient manner. The key idea behind the metamodel approach is to approximate the simulation based objective function ($F(d)$) with a convex and differentiable analytical function ($F^A(d)$), known as the \textit{metamodel}. The metamodel optimization problem, which can be solved efficiently with standard solvers, is then solved sequentially to produce candidate solutions that are evaluated in 
%
For this we utilize the metamodel algorithm proposed in \cite{osorio2019high}. The core idea is to approximate the simulation-based objective function $F(d)$ with a convex and differentiable analytical function $M(d)$ known as the metamodel. The algorithm solves the metamodel optimization problem iteratively to produce candidate solutions that are then evaluated in simulation. In every iteration, the quality of the metamodel is improved by fitting a correction term based on the accumulated simulation evaluations. At the beginning of each iteration, the best OD vector encountered thus far is used as an initial guess for the metamodel optimization sub-problem. We give details of the metamodel problem below and refer the reader to \cite{osorio2019high} for details of the full iterative approach for solving the SO problem. 

% Consider the following notation: 
% \begin{align*}
%     k: & \quad \text{The SO algorithm iteration}\\
%     \beta^k: & \quad \text{Parameter vector of the functional component for the travel time term}\\
%     \alpha^k: & \quad \text{Parameter vector of the functional component for the regularization term}\\
%      \lambda_i, k_i, v_i: & \quad \text{Demand, density, and (space-mean) velocity of segment $i \in \mathcal{I}$}\\
%      l_i, n_i, v_i^{\max}: & \quad \text{Length, number of lanes, and speed limit of segment $i \in \mathcal{I}$}\\
%      v^{\min}, \alpha_1, \alpha_2: & \quad \text{Minimum velocity and fundamental diagram parameters common for all segments}\\
%       \kappa, k^{\text{jam}}, k^{\text{crit}}: & \quad \text{Density scalling factor, jam, and critical densities common for all segments}\\
%       \mathcal{I}_p: & \quad \text{Set of segments in path $p$}\\
%       A: & \quad \text{Assignment matrix that maps OD demand to segment demand}
% \end{align*}
% At every iteration $k$, the metamodel optimization problem is formulated as follows:

Let \(\mathcal{I}_{all}\) denote the set of all segments in the traffic network and $\mathcal{I}_p$ the set of segments comprising path $p$. For segment \( i \in \mathcal{I}_{all} \), we define its length \( l_i \), number of lanes \( n_i \), and speed limit \( v_i^{\max} \). Additionally, \( \lambda_i \), \( k_i \), and \( v_i \) represent analytical demand, density, and space-mean velocity, respectively. For simplicity, we assume all segments share common parameters: minimum velocity \( v^{\min} \), jam density \( k^{\text{jam}} \), critical density \( k^{\text{crit}} \), density scaling factor \( \kappa \), and fundamental diagram parameters, \( \gamma_1 \) and \( \gamma_2 \). $y_p^A$ denotes the analytical travel time of path $p \in \mathcal{P}$. 
At each iteration \( k \), we approximate each term of the simulation-based objective function by an analytical component $f^A(d)$ and a functional component $\phi(d)$, which acts as a correction term for enhanced accuracy. For each functional component, we fit a parameter set: \( \alpha^k \) for the travel time term and \( \beta^k \) for the regularization term. We can now define the metamodel optimization problem for iteration \( k \) as the following convex problem which can be solved efficiently using standard optimization solvers. 
%
%At each iteration \( k \), we fit two sets of parameters: \( \alpha^k \) for the functional component of the travel time term and \( \beta^k \) for the functional component of the regularization term. We can now define the metamodel optimization problem at iteration \( k \) as follows:
\vspace{-15pt}
\begin{align*}
    \min_{0 \leq d \leq d_{max}} & \quad M_k(d; \beta^k, \alpha^k) = w_1 \left(\beta_{0}^k f_1^A(d) + \phi_1(d; \beta^k) \right) + w_2 \left(\alpha_{0}^k f_2^A(d) + \phi_2(d; \alpha^k) \right)\\
    \text{s.t.} & \quad f_1^A(d) = \frac{1}{|\mathcal{P}|} \sum_{p\in \mathcal{P}} \left( y_p^{\text{GT}} - y^A_p \right)^2, \quad 
    \quad f_2^A(d) = \frac{1}{|\mathcal{I}|} \sum_{i\in \mathcal{I}} \left( \frac{\lambda_i}{x_i^{\text{GT}}} - \frac{1}{\mathcal{|I|}}\sum_{j \in \mathcal{I}} \frac{\lambda_j}{x_j^{\text{GT}}} \right)^2, \\
    & \quad \lambda = Ad, \quad \quad k_i = \frac{\kappa k^{\text{jam}}}{n_i} \lambda_i, \quad \forall i \in \mathcal{I}_{all} \\
    & \quad v_i = v^{\min} + (v_i^{\max} - v^{\min}) \left(1 - \left( \frac{\max(k_i, k^{\text{crit}}) - k^{\text{crit}}}{k^{\text{jam}}}\right)^{\gamma_1}\right)^{\gamma_2}, \quad \forall i \in \mathcal{I}_{all}\\
    & \quad y_{p}^A = \sum_{i\in \mathcal{I}_{p}} \frac{l_i}{v_i}, \quad \forall p \in \mathcal{P}\\
    & \quad \phi_1(d; \beta^k) = \beta_1^k + \sum_{z\in \mathcal{Z}} \beta_{z+1}^k d_z, \quad \quad \phi_2(d; \alpha^k) = \alpha_1^k + \sum_{z\in \mathcal{Z}} \alpha_{z+1}^k d_z.
\end{align*}
%
%Each term in the simulation-based objective function is approximated by an analytical component $f^A(d)$ and a functional component $\phi(d)$ that serves as a correction term for improved accuracy. 
%
%This metamodel optimization problem can be solved efficiently using standard convex optimization solvers.  

\vspace{-0.4cm}
\section{Seattle case study} 
\vspace{-0.3cm}
We use the highway network of the Seattle metropolitan area as a case study. The network has 1,820 highway segments and 305 ramp-to-ramp OD pairs. We synthesize three scenarios resembling different congestion levels: low, medium, and high with total demand of $20,000$, $35,000$, and $50,000$, respectively. We generate and simulate the ground truth demand matrices for each scenario. 
Ground truth travel time and count measurements are obtained by sampling $15\%$ of simulated trips and aggregating travel times at the path level and counts at the segment level. We note that this sampling method results in some variability in the penetration rate across ODs (and across segments).

% \begin{table}[h!]
% \footnotesize
%  \centering
%  \begin{tabular}{>{\bfseries}l c c c}
%  \toprule
%  & High congestion & Medium congestion & Low congestion \\
%  \midrule
%  Total Demand & 50000 & 35000 & 20000 \\
%  Minimum OD demand & 10 & 10 & 10 \\
%  Maximum OD demand & 1230 & 961 & 457 \\
%  \bottomrule
%  \end{tabular}
%  \caption{Demand levels for the different scenarios}
%  \label{tab:congestion_demand}
% \end{table}

We compare the proposed formulation to the baseline metamodel approach without regularization in \cite{chao2019traffic}. For every iteration, we evaluate the solution by calculating the normalized root mean squared error (nRMSE) of (1) the demand vector, (2) the path travel times, and (3) the segment counts, averaged over five simulation rollouts, respectively. We use SUMO software \cite{lopez2018microscopic} to model the network and perform all simulations. Each scenario is initialized with a randomized OD vector and the two weights parameters $w_1$ and $w_2$ are selected for each scenario to ensure the objective function terms are of similar magnitude.

\vspace{-0.3 cm}
\subsection{Results} 
\vspace{-0.3 cm}
% In \cref{fig:epochs}, we report each of the three metrics for both the baseline and the regularized metamodel solver at each iteration. Across all scenarios,  the baseline solver demonstrates better performance in terms of travel time, the introduction of the regularization term leads to significant improvements in segment counts and demand fit. Notably, the baseline approach actually deteriorates the estimation of demand, despite its low error in travel time. This highlights a critical drawback of such calibration methods: different demand vectors can yield similar travel time measurements, obscuring the true underlying patterns. The proposed regularization effectively addresses this issue. Through this use case, we illustrate its ability to better reproduce the spatial distribution of the true demand that generated the measurements, leveraging the sample counts across segments. This is further illustrated in

\begin{figure}[t]
    \centering
    \begin{subfigure}[b]{0.45\textwidth}
        \centering
        \includegraphics[width=0.9\linewidth]{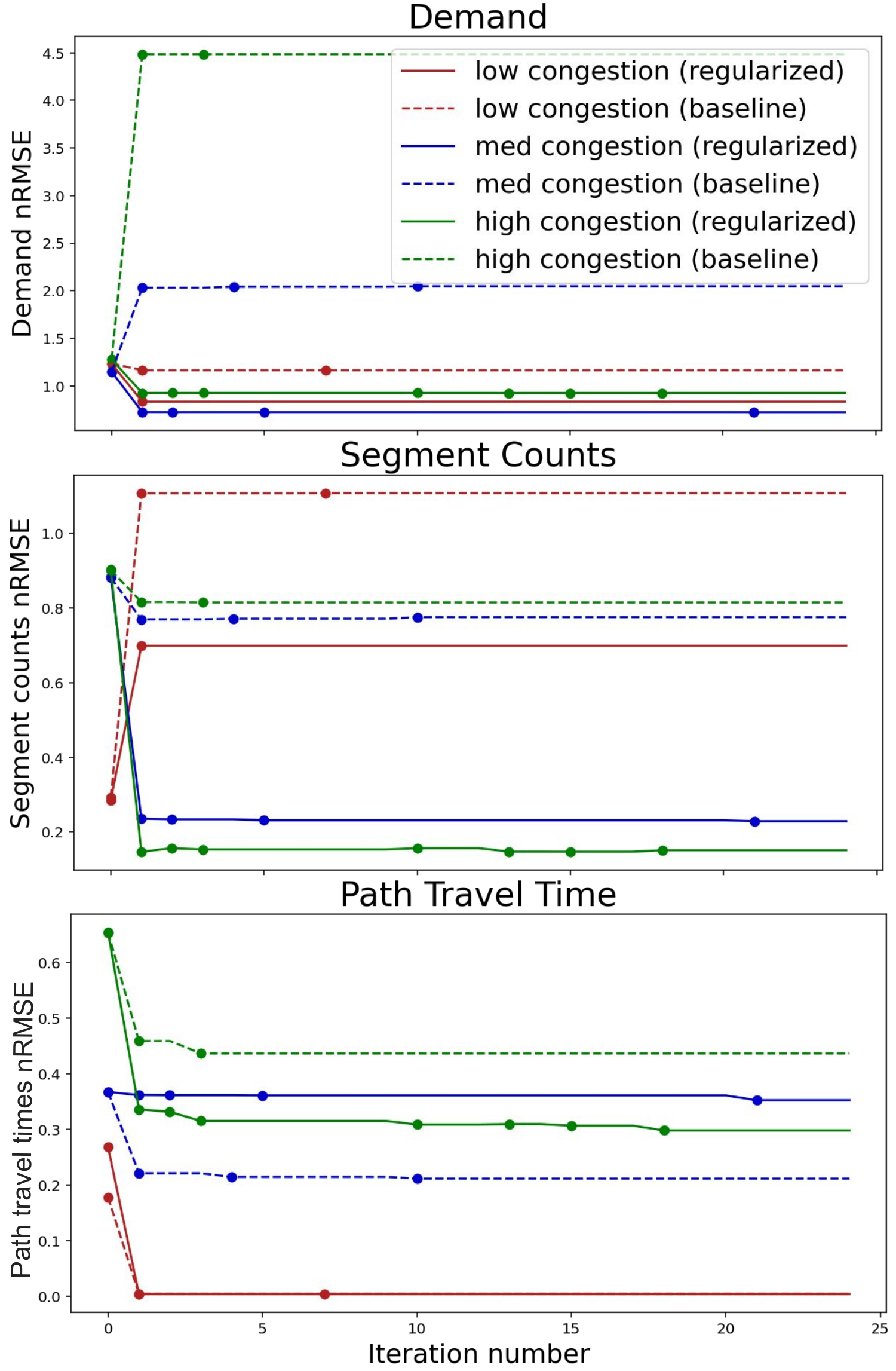}
        \caption{The performance of the calibrated demand across the three metrics at each iteration of the algorithm}
        \label{fig:subplot1}
    \end{subfigure}
    \hfill
    \begin{subfigure}[b]{0.4\textwidth}
        \centering
        \includegraphics[width=0.9\linewidth]{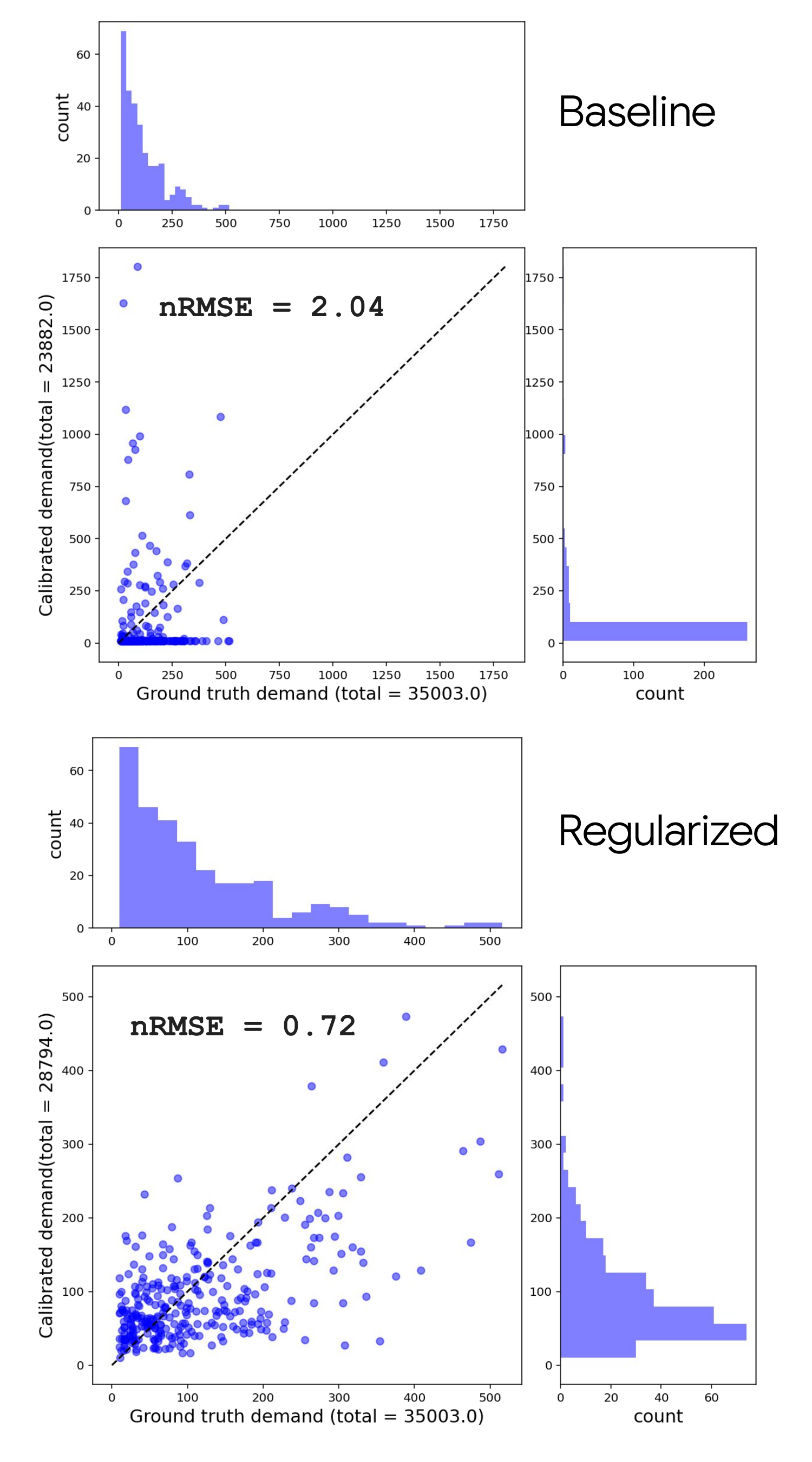}
        \caption{Scatter plot of the ground truth vs. calibrated demand vector (medium congestion).}
        \label{fig:subplot2}
    \end{subfigure}
    \caption{Comparison of the baseline and proposed regularized solver performance across different metrics and scenarios}
\end{figure}

In \cref{fig:subplot1}, we report the three metrics for the baseline and the regularized metamodel solvers at each iteration. While the baseline solver shows better performance in terms of path travel time for the medium and low congestion scenarios, regularization significantly improves the segment counts and demand fit for these scenarios. In the high congestion scenario, the regularized solution outperforms the baseline in all metrics. 
%The relatively higher error in estimating the demand of the high congestion scenario might be due to the large difference between the true demand and the flows realized in the network.
Notably, the baseline solver may even worsen the estimation of demand, to achieve lower path travel time error. This reveals a critical limitation of such calibration methods: different demand vectors can yield similar path travel time measurements, obscuring the true underlying demand patterns. The proposed regularization effectively addresses this issue. Through this use case, we illustrate its ability to better reproduce the spatial distribution of the true demand by leveraging the sample counts across segments. This is further illustrated for the medium demand scenario in \cref{fig:subplot2}. 
%where we show scatter plots of both solvers solutions against the ground truth demand.
%Through this use case, we illustrate its ability to better reproduce the spatial distribution of the true demand by leveraging the sample counts across segments. 

\vspace{-0.4 cm}
\section{Discussion} 
\vspace{-0.3 cm}
This work presents a novel formulation for demand calibration, introducing a new segment count regularization method to tackle the underdetermination challenge.
Our approach offers significant improvements in recovering the true demand vector, achieving up to a $65\%$ decrease in terms of nRMSE compared to the unregularized solution, with only a slight degradation in the travel time performance. The proposed formulation exhibits generalizability, enabling its application to a wide range of calibration objectives, leveraging diverse field measurement data. 
%In this work, we propose a novel formulation for simulation-based demand calibration. Our approach introduces a new method for regularizing the optimization problem to address the challenge of underdetermination. We observe significant improvements in recovering the true demand vector, achieving up to a $65\%$ decrease in the nRMSE compared to the unregularized calibrated vector, with only a slight degradation in travel time performance. Furthermore, this regularization formulation can be applied to any other calibration objective, such as segment speeds.

%  \begin{figure}[h!]
%      \centering
%      \includegraphics[width = 0.4\linewidth]{figs/epoch_lines.png}
%      \caption{The performance of the calibrated demand across the three metrics at each iteration of the algorithm}
%      \label{fig:epochs}
%  \end{figure}

%   \begin{figure}
%      \centering
%      \includegraphics[width = 0.8 \linewidth]{figs/scatter_med.png}
%      \caption{Scatter plot of the ground truth vs. calibrated demand vector for the baseline and the proposed solvers (medium congestion scenario).}
%      \label{fig:scatter_med}
%  \end{figure}
\vspace{-0.4 cm}

\begin{small}
\begin{sloppypar} 
\bibliographystyle{authordate1} 

\setlength{\bibsep}{0pt}
\bibliography{refs}

\begin{thebibliography}{}

\bibitem[\protect\citename{Barcelo {\em et~al.}, }2012]{barcelo2012dynamic}
Barcelo, Jaume, Montero, L{\'\i}dia, Bullejos, Manuel, Serch, Oriol, \&
  Carmona, C. 2012.
\newblock Dynamic OD matrix estimation exploiting bluetooth data in urban
  networks.
\newblock {\em Recent researches in automatic control and electronics},
  116--121.

\bibitem[\protect\citename{Ben-Akiva {\em et~al.}, }2012]{ben2012dynamic}
Ben-Akiva, Moshe~E, Gao, Song, Wei, Zheng, \& Wen, Yang. 2012.
\newblock A dynamic traffic assignment model for highly congested urban
  networks.
\newblock {\em Transportation research part C: Emerging Technologies}, {\bf
  24}, 62--82.

\bibitem[\protect\citename{Chiappone {\em et~al.}, }2016]{chiappone2016traffic}
Chiappone, Sandro, Giuffr{\`e}, Orazio, Gran{\`a}, Anna, Mauro, Raffaele, \&
  Sferlazza, Antonino. 2016.
\newblock Traffic simulation models calibration using speed--density
  relationship: An automated procedure based on genetic algorithm.
\newblock {\em Expert Systems with Applications}, {\bf 44}, 147--155.

\bibitem[\protect\citename{Iqbal {\em et~al.}, }2014]{iqbal2014development}
Iqbal, Md~Shahadat, Choudhury, Charisma~F, Wang, Pu, \& Gonz{\'a}lez, Marta~C.
  2014.
\newblock Development of origin--destination matrices using mobile phone call
  data.
\newblock {\em Transportation Research Part C: Emerging Technologies}, {\bf
  40}, 63--74.

\bibitem[\protect\citename{Lopez {\em et~al.}, }2018]{lopez2018microscopic}
Lopez, Pablo~Alvarez, Behrisch, Michael, Bieker-Walz, Laura, Erdmann, Jakob,
  Fl{\"o}tter{\"o}d, Yun-Pang, Hilbrich, Robert, L{\"u}cken, Leonhard, Rummel,
  Johannes, Wagner, Peter, \& Wie{\ss}ner, Evamarie. 2018.
\newblock Microscopic traffic simulation using sumo.
\newblock {\em In:} {\em 21st ITSC}.
\newblock IEEE.

\bibitem[\protect\citename{Mo {\em et~al.}, }2020]{mo2020estimating}
Mo, Baichuan, Li, Ruimin, \& Dai, Jingchen. 2020.
\newblock Estimating dynamic origin--destination demand: A hybrid framework
  using license plate recognition data.
\newblock {\em Computer-Aided Civil and Infrastructure Engineering}, {\bf
  35}(7), 734--752.

\bibitem[\protect\citename{Osorio, }2019]{osorio2019high}
Osorio, Carolina. 2019.
\newblock High-dimensional offline origin-destination (OD) demand calibration
  for stochastic traffic simulators of large-scale road networks.
\newblock {\em Transportation Research Part B: Methodological}, {\bf 124},
  18--43.

\bibitem[\protect\citename{Zhang {\em et~al.}, }2024]{chao2019traffic}
Zhang, Chao, Li, Yechen, Arora, Neha, Pierce, Damien, \& Osorio, Carolina.
  2024.
\newblock Traffic Simulations: Multi-City Calibration of Metropolitan Highway
  Networks.
\newblock {\em In:} {\em 27th IEEE international conference on intelligent
  transportation systems (ITSC)}.

\end{thebibliography}

\end{sloppypar}
\end{small}

%%%%%%%%%%%%%%%%%%%%%%%%%%%%%%%%%%%%%%%%%%%%%%%%%%%%%%%%%%%%%
%% APPENDICES
%%%%%%%%%%%%%%%%%%%%%%%%%%%%%%%%%%%%%%%%%%%%%%%%%%%%%%%%%%%%%

\end{document}